\documentclass[a4paper,fleqn,12pt]{article}
\usepackage{graphicx}
\usepackage[small]{subfigure,epsfig}
\usepackage{cite}        
\usepackage{amssymb,amsmath,latexsym,enumerate}
\usepackage{color}
\numberwithin{equation}{section}

\newcommand{\PDfrac}[3][{}]{ \frac{\partial^{#1}#2}{\partial#3^{#1}} }

\begin{document}

\title{Extended equation for description of nonlinear waves in liquid with gas bubbles }

\author{Nikolai A. Kudryashov, \and Dmitry I. Sinelshchikov}

\date{Department of Applied Mathematics, National Research Nuclear University
MEPHI, 31 Kashirskoe Shosse, 115409 Moscow, Russian Federation}

\maketitle

\begin{abstract}
Nonlinear waves in a liquid with gas bubbles are studied. Higher order terms with respect to the small parameter are taken into account in the derivation of the equation for nonlinear  waves. A nonlinear differential equation is derived for long weakly nonlinear waves taking into consideration liquid viscosity, inter--phase heat transfer and surface tension. Additional conditions for the parameters of the equation are determined for integrability of the mathematical model. The  transformation for linearization of the nonlinear equation is presented too. Some exact solutions of the nonlinear equation are found for integrable and non--integrable cases. The nonlinear waves described by the nonlinear equation are numerically investigated.
\end{abstract}

\section{Introduction}
It is known that a liquid with gas bubbles is often observed in nature, medicine and industry \cite{Nakor'akov,Nigmatulin,Goldberg}. One of the important problems is the investigation of nonlinear waves in bubble--liquid mixtures.

Study  of nonlinear waves processes in a bubble-liquid mixture was first carried out in \cite{Wijngaarden, Wijngaarden1972, Nakor'akov_1972}. In these works the Burgers, the Korteweg--de Vries and the Burgers--Korteweg--de Vries equations were derived for description of weakly nonlinear waves in the one--dimensional case. The nonlinear waves in a bubbly liquid with an allowance for the interphase heat transfer were considered in \cite{Prosperetti1991, Ogonyan00, Kudryashov2010, Kudryashov2010A}. The nonlinear evolution equation for weakly nonlinear waves in a liquid with gas bubbles in the three--dimensional case were obtained in \cite{Kudryashov2012}.

However only the first order terms with respect to the small parameter were taken into account in the derivation of nonlinear evolution equations in the above mentioned works. It is known that the application of higher order terms in the state equation allows us to obtain a more exact description of nonlinear waves \cite{Depassier1992,Sibgatullin,Holm2003,Holm2004,Randruut2010,Depassier2012}. Thus the important problem is to derive a nonlinear differential equation for description of long weakly nonlinear waves in a bubbly liquid taking into consideration the second order terms in the asymptotic expansion.

The aim of this work is to derive the nonlinear differential equation for description of long waves in a liquid with gas bubbles taking into account higher order terms in the asymptotic expansion. In the model of nonlinear waves we include the surface tension, the liquid viscosity and the inter--phase heat exchange in the quasi-isothermal regime.

This work is organized as follows. In section 2 we present basic equations for the description of nonlinear waves. In section 3 we derive the extended version of the equation for the description of nonlinear waves in a liquid with gas bubbles taking into consideration the second order terms with respect to the small parameter. In section 4 we apply the Painlev\`{e} test for the new nonlinear equation. We show that this equation is integrable under additional  conditions on parameters. We obtain some exact solutions for the integrable case of the nonlinear equation in section 5. In section 6 we consider the traveling wave solutions for the non--integrable case of the nonlinear equation. We also discuss connections of the improved equation with other nonlinear differential equations. We show that the nonlinear equation under some conditions on parameters is equivalent to the equation obtained in \cite{Kawahara} for the description of the ion-acoustic waves in plasma. In section 7 we present the results of the numerical simulations for nonlinear waves governed by the nonlinear equation.

\section{System of equations for description of waves in a liquid with gas bubbles}

Let us suppose that a liquid containing gas bubbles is a homogeneous medium and has an average pressure \cite{Nakor'akov,Nigmatulin}. We do not take into consideration formation, destruction, interaction and coalescence of bubbles. Let us assume that all gas bubbles are spherical, have the same size and the amount of bubbles in the mass unit is the constant $N$. We take into consideration the heat transfer between a gas in bubbles and a liquid in the nearly isothermal approximation \cite{Prosperetti1991}. In this approximation it is supposed that the temperature of the liquid is not changed and is equal to the temperature of the mixture in the unperturbed state \cite{Prosperetti1991}. We consider the influence of the liquid viscosity only at the interphase boundary. We also assume that the problem is one--dimensional.  Under these assumptions the following closed system of equations may be used for a description of nonlinear waves in the liquid with gas bubbles \cite{Nakor'akov,Nigmatulin,Prosperetti1991}
\begin{equation}
\rho_{\tau}+(\rho\,u)_{\xi}=0, \hfill
\label{pre_main_main_dim_system_1}
\end{equation}
\begin{equation}
\rho\left(u_{\tau}+u u_{x}\right)+P_{\xi}=0,
\label{pre_main_main_dim_system_2}
\end{equation}
\begin{equation}
\rho_{l}\left(R R_{\tau\tau}+\frac{3}{2} R_{\tau}^{2}+\frac{4\nu_{l}}{3R}R_{\tau}\right)=P_{g}-P-\frac{2\sigma}{R},
\label{pre_main_main_dim_system_3}
\end{equation}
\begin{equation}
\begin{gathered}
P_{g}=P_{g,0}\left(\frac{R_{0}}{R}\right)^{3}\Bigg\{1-\frac{(\tilde{\gamma}-1)R_{0}^{3}}{5\tilde{\gamma}\chi}\,\,\frac{R_{\tau}}{R^{2}}-
\frac{1(\tilde{\gamma}-1)^{2}R_{0}^{6}}{525\tilde{\gamma}^{2}\chi^{2}} \times   \\ \times \Bigg[(2+15K_{0}^{'})\,\,\frac{R_{\tau}^{2}}{R^{4}}
+\frac{12\tilde{\gamma}-7}{3(\tilde{\gamma}-1)R}\left(\frac{R_{\tau\tau}}{R^{2}}-\frac{2R_{\tau}^{2}}{R^{3}}\right)\Bigg]\Bigg\}
\label{pre_main_main_dim_system_4}
\end{gathered}
\end{equation}
\begin{equation}
\rho=(1-\phi)\,\rho_{l}+\phi\,\rho_{g},
\label{pre_main_main_dim_system_5}
\end{equation}
\begin{equation}
\phi=V\,\rho, \quad V=\frac{4}{3}\,\pi\,R^{3}\,N,
\label{pre_main_main_dim_system_6}
\end{equation}
where $\xi$ is cartesian coordinate, $\tau$ is time, $\rho(\xi,\tau)$ is the density of the bubble--liquid mixture, $P(\xi,\tau)$ is the pressure of the mixture, $u(\xi,\tau)$ is the velocity of the mixture, $R=R(\xi,\tau)$ is the bubble radius, $\rho_{l}, \rho_{g}(\xi,\tau)$ are densities of the liquid and the gas respectively, $P_{g}(\xi,\tau)$ is the pressure of the gas, $P_{g,0}$ and $R_{0}$ are the pressure of the gas and the radius of bubbles in the unperturbed state, $\sigma$ is the surface tension, $\nu_{l}$ is the kinematic viscosity of the liquid, $\chi$ is the thermal diffusivity of the gas, $K$ is the thermal conductivity of the gas and $K_{0}^{'}=d K/d T$ at $T=T_{0}$, $\phi$ is the volume gas content, $V$ is the gas volume in the unit mass of the mixture, $\tilde{\gamma}$ is the ratio of the specific heats.

We note that Eq.\eqref{pre_main_main_dim_system_4} was derived in \cite{Prosperetti1991} under the assumption that the thermal penetration length is large compared with the radius of the bubble.

We assume that the pressure and the density of the bubble--liquid mixture are constants in the unperturbed state. We also assume that all bubbles have the same radius and uniformly distributed in the liquid in the unperturbed state.

Assuming that the volume gas content is small, $\phi \ll 1$,   from Eqs. \eqref{pre_main_main_dim_system_5} and \eqref{pre_main_main_dim_system_6} we have the following equation
\begin{equation}
\rho=\frac{\rho_{l}}{1+\rho_{l}\,V}\,\,.
\label{eq: density}
\end{equation}
Let us suppose that deviation of the mixture density is small
\begin{equation}
 \rho(\xi,\tau) = \rho_{0} + \tilde{\rho}(\xi,\tau),\:\   \quad
  \rho_{0}=\mbox{const},\quad  ||\tilde{\rho}||<<\rho_{0},
  \label{eq: rel1}
\end{equation}
where $\rho_{0}$ is the density of the bubble--liquid mixture in the unperturbed state.

Taking into account Eq.\eqref{eq: rel1} we obtain from Eq.\eqref{eq: density}
\begin{equation}
\begin{gathered}
 R\simeq R_{0}-\tilde{\mu}\tilde{\rho}+\tilde{\mu}_{1}\tilde{\rho}^{2},      \\
R_{0}^{3}=\frac{3}{4\pi N}\left(\frac{1}{\rho_{0}}-\frac{1}{\rho_{l}}\right), \quad \tilde{\mu}=\frac{R_{0}}{3\,\rho_{0}^{2}\,V_{0}}, \quad
\tilde{\mu}_{1}=\frac{R_{0}(3\,\rho_{0}V_{0}-1)}{9\,\rho_{0}^{4}\,V_{0}^{2}},  \\
V_{0}=\frac{4}{3}\pi N R_{0}^{3}\,.
  \label{eq:R_equation}
\end{gathered}
\end{equation}

Substituting Eq.\eqref{eq:R_equation} into Eqs.\eqref{pre_main_main_dim_system_1}--\eqref{pre_main_main_dim_system_4} and taking into account \eqref{eq: rel1} we have
\begin{equation}
\begin{gathered}
\tilde{\rho}_{\tau}+\rho_{0}u_{\xi}+(\tilde{\rho} u)_{\xi}=0, \hfill \vspace{0.1cm} \\
(\rho_{0}+\tilde{\rho})\left(u_{\tau}+u u_{\xi}\right)+P_{\xi}=0, \hfill \vspace{0.1cm} \\
P=P_{0}-\frac{2\sigma}{R_{0}}+\left(\frac{3\,\tilde{\mu}\,P_{0}}{R_{0}}-\frac{2\sigma\tilde{\mu}}
{R_{0}^{2}}\right)\,\tilde{\rho}+\hfill \vspace{0.1cm}  \\
+\left(\frac{[6\tilde{\mu}^{2}-3\,\tilde{\mu}_{1}\,R_{0}]P_{0}}{R_{0}^{2}}-\frac{2\sigma
(\tilde{\mu}^{2}-\tilde{\mu}_{1}R_{0})}{R_{0}^{3}}\right)\,\tilde{\rho}^{2}+\hfill \vspace{0.1cm}\\+
\left(\rho_{l}\,\tilde{\mu}\,R_{0}+\frac{\lambda_{2}\gamma_{2}\tilde{\mu} P_{0}}{R_{0}^{3}}\right)\tilde{\rho}_{\tau\tau}+\hfill \vspace{0.1cm} \\
+\Big[\frac{2\lambda_{2}\gamma_{2}P_{0}(3\tilde{\mu}^{2}-\tilde{\mu}_{1}R_{0})}{R_{0}^{4}}-
\rho_{l}(2\tilde{\mu}_{1}\,R_{0}+\tilde{\mu}^{2})\Big]\tilde{\rho}\,\tilde{\rho}_{\tau\tau}
-\hfill \vspace{0.1cm}\\
-\left[\frac{\lambda_{2}P_{0}[(\gamma_{1}-2\gamma_{2})\tilde{\mu}^{2}+2\tilde{\mu}_{1}\gamma_{2}R_{0}]}{R_{0}^{4}}+
\rho_{l}\left(2\tilde{\mu}_{1}\,R_{0}+\frac{3\tilde{\mu}^{2}}{2}\right)\right]\,\tilde{\rho}_{\tau}^{2}+
\hfill \vspace{0.1cm}  \\
+\Big(\frac{\lambda_{1}\tilde{\mu} P_{0}}{R_{0}^{2}}+\frac{4\nu_{l}\tilde{\mu}\,\rho_{l}}{3R_{0}}\Big)\,\tilde{\rho}_{\tau}+\hfill \vspace{0.1cm}  \\
+\Big[\frac{\lambda_{1}(5\tilde{\mu}^{2}-2\tilde{\mu}_{1}R_{0})P_{0}}{R_{0}^{3}}+
\frac{(4\nu_{l}\tilde{\mu}^{2}-8\nu_{l}\tilde{\mu}_{1}R_{0})\rho_{l}}{3R_{0}^{2}}\Big]\,\tilde{\rho}\,
\tilde{\rho}_{\tau}, \hfill
 \label{dim_system}
\end{gathered}
\end{equation}
where $P_{0}$ is the pressure of gas in bubbles in the unperturbed state. We used the following notations in \eqref{dim_system}:
\begin{equation}
\begin{gathered}
\lambda_{1}=\frac{(\tilde{\gamma}-1)R_{0}^{3}}{5\tilde{\gamma}\chi},\quad
\lambda_{2}=\frac{(\tilde{\gamma}-1)^{2}R_{0}^{6}}{525\tilde{\gamma}^{2}\chi^{2}},\vspace{0.1cm}\\
\gamma_{1}=(2+15 K_{0}^{'}),\quad \gamma_{2}=\frac{12\tilde{\gamma}-7}{3(\tilde{\gamma}-1)}.
\end{gathered}
\end{equation}

Linearizing Eqs.\eqref{dim_system} and assuming that $P$ is proportional to $\tilde{\rho}$ we obtain the linear wave equation
\begin{equation}
\tilde{\rho}_{\tau\tau}=c_{0}^{2} \, \tilde{\rho}_{\xi\xi}, \quad c_{0}^{2}=\frac{3\tilde{\mu} P_{0}}{R_{0}}-\frac{2\sigma\tilde{\mu}}{R_{0}^{2}}\,.
\label{wave_equation}
\end{equation}
From \eqref{wave_equation} we see that $c_{0}$ is the speed of linear waves.

Let us introduce the following dimensionless variables
\begin{equation}
\xi = L\, \xi^{'},  \quad   \tau= \frac{ L }{ c_{0} }\, \tau', \quad  u = c_{0}\,u^{'}, \\
  \tilde{\rho}=\rho_{0} \tilde{\rho}^{'}, \quad P = P_0\, P'+P_0-\frac{2\sigma}{R_{0}},
  \label{eq: non-dim_subst}
\end{equation}
where $P_{0}-2\sigma/R_{0}$ is the pressure of the mixture in the unperturbed state. The quantities $L$ and $\tau_{*}=L/c_{0}$ are the characteristic length scale and the characteristic time of our problem \cite{Nigmatulin}.

Using the dimensionless variables we can reduce \eqref{dim_system} to the following system of equations (the primes are omitted)
\begin{equation}
\begin{gathered}
\tilde{\rho}_{\tau}+u_{\xi}+(\tilde{\rho} u)_{\xi}=0, \hfill \vspace{0.1cm} \\
(1+\tilde{\rho})\left( u_{\tau}+u u_{\xi}\right)+
\frac{1}{\tilde{\alpha}} P_{\xi}=0,\hfill \vspace{0.1cm} \\
P=\tilde{\alpha}\tilde{\rho}+\tilde{\alpha}_{1}\tilde{\rho}^{2}+
\beta_{1}\tilde{\rho}_{\tau\tau}-\beta_{2}\tilde{\rho}\,\tilde{\rho}_{\tau\tau}-
\beta_{3}\tilde{\rho}_{\tau}^{2}+
\varkappa\tilde{\rho}_{\tau}+\varkappa_{1}\tilde{\rho}\tilde{\rho}_{\tau}, \hfill
\label{main_nondim_system}
\end{gathered}
\end{equation}
where
\begin{equation}
\begin{gathered}
    \tilde{\alpha} = \frac{3\tilde{\mu}\rho_{0}}{R_{0}}-\frac{2\sigma\tilde{\mu}\rho_{0}}{R_{0}^{2}P_{0}},\quad
    \beta_{1}=\frac{\rho_{l} \tilde{\mu} R_{0} c_{0}^{2} \rho_{0}}{P_{0}\,L^{2}}+\frac{\gamma_{2}\rho_{0}\tilde{\mu}(\tilde{\gamma}-1)^{2}}{525R_{0}\tilde{\gamma}^{2}D^{2}}, \vspace{0.1cm} \\
    \tilde{\alpha}_{1}=\frac{[(6\tilde{\mu}^{2}-3\,\tilde{\mu}_{1}\,R_{0}]\rho_{0}^{2}}{R_{0}^{2}}
    -\frac{2\sigma(\tilde{\mu}^{2}-\tilde{\mu}_{1}R_{0})\rho_{0}^{2}}{P_{0}R_{0}^{3}}, \vspace{0.1cm} \\
    \beta_{2} =\frac{2(\tilde{\gamma}-1)^{2}(\tilde{\mu}_{1}R_{0}-3\tilde{\mu}^{2})\rho_{0}^{2}}{525\tilde{\gamma}^{2}R_{0}^{2}D^{2}}+ \frac{\rho_{l}(2 \tilde{\mu}_{1} R_{0}+\tilde{\mu}^{2}) c_{0}^{2} \rho_{0}^{2}}{P_{0}\,L^{2}}, \vspace{0.1cm} \\
    \beta_{3} = \frac{(\tilde{\gamma}-1)^{2}[(\gamma_{1}-2\gamma_{2})\tilde{\mu}^{2}+2\tilde{\mu}_{1}\gamma_{2}R_{0}]\rho_{0}^{2}}{525\tilde{\gamma}^{2}R_{0}^{2}D^{2}}+
    \frac{\rho_{l}(4\tilde{\mu}_{1}R_{0}\tilde{\mu}^{2}+3\tilde{\mu}^{2}) \rho_{0}^{2} c_{0}^{2}}{2P_{0}\,L^{2}}, \vspace{0.1cm} \\
    \varkappa=\frac{(\tilde{\gamma}-1)\rho_{0}\tilde{\mu}}{5\tilde{\gamma} R_{0}D}+\frac{4\nu_{l}\tilde{\mu}\,\rho_{l}\rho_{0}c_{0}}{3R_{0}P_{0}L},\vspace{0.1cm} \\
    \varkappa_{1}=\frac{(\tilde{\gamma}-1)(5\tilde{\mu}^{2}-2\tilde{\mu}_{1}R_{0})\rho_{0}^{2}}{5\tilde{\gamma} R_{0}^{2} D}+\frac{(4\nu_{l}\tilde{\mu}^{2}-8\nu_{l}\tilde{\mu}_{1}R_{0})\rho_{l}\rho_{0}^{2}c_{0}}
    {3R_{0}^{2}P_{0}L}\,.
\end{gathered}
  \label{eq: non-dim_parameters}
\end{equation}
We used in \eqref{eq: non-dim_parameters} the notation
\begin{equation}
D=\frac{\chi }{\omega R_{0}^{2}},\,\, \mbox{where}\,\,\,  \omega=\frac{c_{0}}{L}\,.
\label{D_parametr}
\end{equation}
The parameter $D$ in \eqref{D_parametr} is the square of the ratio of the thermal penetration length to the radius of bubbles in the unperturbed state \cite{Prosperetti1991}. In the case of the isothermal behavior of bubbles, parameter $D$ tends to infinity and system of equations \eqref{main_nondim_system} is reduced to the system that was used in some works (e.g. see \cite{Wijngaarden1972,Nakor'akov_1972}).

\section{Extended equation for nonlinear waves in a liquid with gas bubbles.}

We use the reductive perturbation method \cite{Kevorkian1985,Taniuti1990,Gardner1969,Leblond2008} for derivation of the extended equation. In accordance with this method we need to introduce the 'slow' variables
\begin{equation}
\begin{gathered}
  x = \varepsilon^m(\xi-\tau), \quad t= \varepsilon^{m+1}\, \tau ,\quad
  m > 0, \quad \varepsilon \ll 1,
  \label{eq: rescaling_coordinates}
    \end{gathered}
\end{equation}
\begin{equation*}
  \frac{\partial}{\partial \xi} = \varepsilon^m \frac{\partial}{\partial x},\quad
  \frac{\partial}{\partial \tau} = \varepsilon^{m+1} \frac{\partial}{\partial t}
  - \varepsilon^m \frac{\partial}{\partial x}\,.
\end{equation*}

Substituting \eqref{eq: rescaling_coordinates} into \eqref{main_nondim_system} and dividing by $\varepsilon^{m}$  each side of the first two equations we obtain the system of equations
\begin{equation}
\begin{gathered}
\varepsilon \tilde{\rho}_{t}-\tilde{\rho}_{x}+u_{x}+(\tilde{\rho} u)_{x}=0,\hfill \\
(1+\tilde{\rho})(\varepsilon u_{t}-u_{x}+u u_{x})+\frac{1}{\tilde{\alpha}} P_{x}=0, \hfill \\
P=\tilde{\alpha} \tilde{\rho}+\tilde{\alpha}_{1} \tilde{\rho}^{2}+
\varepsilon^{2m+2}\beta_{1} \tilde{\rho}_{tt}-\varepsilon^{2m+1}2 \beta_{1} \tilde{\rho}_{tx}+\varepsilon^{2m}\beta_{1} \tilde{\rho}_{xx}-\varepsilon^{2m+2}\beta_{2} \tilde{\rho} \tilde{\rho}_{tt}+\hfill \\
+\varepsilon^{2m+1}2 \beta_{2}\tilde{\rho} \tilde{\rho}_{tx}-\varepsilon^{2m}\beta_{2}\tilde{\rho}\tilde{\rho}_{xx}
-\varepsilon^{2m+2}\beta_{3} \tilde{\rho}_{t}^{2}
+\varepsilon^{2m+1}2 \beta_{3}\tilde{\rho}_{x} \tilde{\rho}_{t}-\varepsilon^{2m}\beta_{3}\tilde{\rho}_{x}^{2}+\hfill \\
+\varepsilon^{m+1}\varkappa \tilde{\rho}_{t}-\varepsilon^{m}\varkappa \tilde{\rho}_{x}
+\varepsilon^{m+1}\varkappa_{1} \tilde{\rho} \tilde{\rho}_{t}-\varepsilon^{m}\varkappa_{1} \tilde{\rho} \tilde{\rho}_{x}\,. \hfill
\label{rescaled_system}
\end{gathered}
\end{equation}

We search for the solutions of \eqref{rescaled_system} in the form of a series in the small parameter $\varepsilon$
\begin{equation}
  \begin{gathered}
    \tilde{\rho}  =\varepsilon \tilde{\rho}_1 + \varepsilon^{2} \tilde{\rho}_2 +\ldots,  \\
    u = \varepsilon u_1 + \varepsilon^{2} u_2 +  \ldots,  \\
    P = \varepsilon P_1  + \varepsilon^{2} P_2   +\ldots \,.
  \end{gathered}
  \label{eq: asymptotic_expansion1}
\end{equation}

Substituting \eqref{eq: asymptotic_expansion1} into \eqref{rescaled_system} and collecting coefficients at order $\varepsilon$ we have
\begin{equation}
  \begin{gathered}
u_{1}(x,t)=\tilde{\rho}_{1}(x,t), \quad P_{1}(x,t)=\tilde{\alpha} \tilde{\rho}_{1}(x,t)\,.
  \end{gathered}
  \label{eq: first_order_relation}
\end{equation}

Substituting \eqref{eq: asymptotic_expansion1} into \eqref{rescaled_system}  and collecting coefficients at order $\varepsilon^{2}$, we obtain
\begin{equation}
\begin{gathered}
\tilde{\rho}_{1t}-\tilde{\rho}_{2x}+u_{2x}+(\tilde{\rho}_{1}u_{1})_{x}=0,\hfill\\
u_{1t}-u_{2x}+u_{1}u_{1x}-\tilde{\rho}_{1}u_{1x}+\frac{1}{\tilde{\alpha}}P_{2x}=0,\hfill\\
P_{2}=\tilde{\alpha} \tilde{\rho}_{2}+\tilde{\alpha}_{1}\tilde{\rho}_{1}^{2}+\varepsilon^{2m-1}\beta_{1}\tilde{\rho}_{1xx}+\varepsilon^{m}\varkappa\tilde{\rho}_{1t}-
\varepsilon^{m-1}\varkappa\tilde{\rho}_{1x}-\varepsilon^{m}\varkappa_{1}\tilde{\rho}_{1}\tilde{\rho}_{1x}-\hfill\\
-\varepsilon^{2m}\beta_{2}\tilde{\rho}_{1}\tilde{\rho}_{1xx}-\varepsilon^{2m}\beta_{3}\tilde{\rho}_{1x}^{2}-\varepsilon^{2m}2 \beta_{1} \tilde{\rho}_{1tx}+\varepsilon^{m+1}\varkappa_{1} \tilde{\rho} \tilde{\rho}_{t}\,. \hfill
\label{eq: second_order_relations}
 \end{gathered}
\end{equation}

Using relations \eqref{eq: first_order_relation} from Eqs.\eqref{eq: second_order_relations} we have the nonlinear differential equation
\begin{equation}
\begin{gathered}
\tilde{\rho}_{1t}+\left(1+\frac{\tilde{\alpha}_{1}}{\tilde{\alpha}}\right)\tilde{\rho}_{1}\tilde{\rho}_{1x}+\varepsilon^{2m-1}\frac{\beta_{1}}{2\tilde{\alpha}}\tilde{\rho}_{1xxx}+
\varepsilon^{m}\frac{\varkappa}{2\tilde{\alpha}}\tilde{\rho}_{1tx}-\varepsilon^{m-1}\frac{\varkappa}{2\tilde{\alpha}}\tilde{\rho}_{1xx}-\hfill \vspace{0.1cm}\\-
\varepsilon^{m}\frac{\varkappa_{1}}{2\tilde{\alpha}}(\tilde{\rho}_{1}\tilde{\rho}_{1x})_{x}
-\varepsilon^{2m}\frac{\beta_{2}}{2\tilde{\alpha}}(\tilde{\rho}_{1}\tilde{\rho}_{1xx})_{x}
-\varepsilon^{2m}\frac{\beta_{3}}{2\tilde{\alpha}}(\tilde{\rho}_{1x}^{2})_{x}-\hfill\vspace{0.1cm}\\
-\varepsilon^{2m} \frac{\beta_{1}}{\tilde{\alpha}} \tilde{\rho}_{1txx}+\varepsilon^{m+1}\frac{\varkappa_{1}}{2\tilde{\alpha}}
(\tilde{\rho} \tilde{\rho}_{t})_{x} =0\,. \hfill
\label{eq: mail_evolution_equation}
 \end{gathered}
\end{equation}

From Eq.\eqref{eq: mail_evolution_equation} we see that at $m=1$ the term with lowest power of $\varepsilon$ corresponds to the term with the second derivative. If we consider the terms with the next power of $\varepsilon$ then we have to add the nonlinear dissipative term $(\rho_{1}\rho_{1x})_{x}$, a dispersive term and the term $\rho_{1tx}$ into the equation. In this case at $m=1$ from Eq.\eqref{eq: mail_evolution_equation} we obtain
\begin{equation}
\tilde{\rho}_{1t}+\left(1+\frac{\tilde{\alpha}_{1}}{\tilde{\alpha}}\right)\tilde{\rho}_{1}\tilde{\rho}_{1x}-\frac{\varkappa}{2\tilde{\alpha}}\tilde{\rho}_{1xx}+\varepsilon\left( \frac{\beta_{1}}{2\tilde{\alpha}}\tilde{\rho}_{1xxx}+\frac{\varkappa}{2\tilde{\alpha}}\tilde{\rho}_{1tx}-
\frac{\varkappa_{1}}{2\tilde{\alpha}}(\tilde{\rho}_{1}\tilde{\rho}_{1x})_{x}\right)=0\,.
\label{eq: mail_evolution_equation1}
\end{equation}

Eq.\eqref{eq: mail_evolution_equation1} extends the Burgers equation \cite{Nakor'akov_1972,Wijngaarden1972} for long weakly nonlinear waves in the liquid containing gas bubbles. The last three terms in Eq.\eqref{eq: mail_evolution_equation1} are the linear dispersive term and linear and nonlinear dissipative terms.

Using relations \eqref{eq: non-dim_parameters} we see that dissipation of nonlinear waves governed by Eq.\eqref{eq: mail_evolution_equation1} depends on the liquid viscosity, the relation between length of thermal penetration, the radius of bubbles and the ratio of specific heats. In the purely isothermal case the dissipation is caused by the liquid viscosity. Otherwise the dissipation is caused by the heat transfer and the liquid viscosity.

The coefficient at the dispersion term in Eq.\eqref{eq: mail_evolution_equation1} depends on the unperturbed pressure, the radius of bubbles and the wave speed in the linear case. There is a `thermal' part of the dispersion coefficient in Eq.\eqref{eq: mail_evolution_equation1} as well.

For convenience of description let us use the following notations in Eq.\eqref{eq: mail_evolution_equation1}
\begin{equation}
\begin{gathered}
\left(1+\frac{\tilde{\alpha}_{1}}{\tilde{\alpha}}\right)=\alpha,\quad \varepsilon \frac{\beta_{1}}{2\tilde{\alpha}} = \beta, \quad \varepsilon\frac{\varkappa}{2\tilde{\alpha}}=\gamma,\\
\frac{\varkappa}{2\tilde{\alpha}}=\mu, \quad \varepsilon \frac{\varkappa_{1}}{2\tilde{\alpha}}=\nu,\\
\tilde{\rho}_{1}=v\,.
\label{notations}
\end{gathered}
\end{equation}

Using \eqref{notations} from Eq.\eqref{eq: mail_evolution_equation1} we have
\begin{equation}
v_{t}+\alpha v v_{x}+\beta v_{xxx}-\mu v_{xx}-\nu (v v_{x})_{x}+\gamma v_{xt}=0\,.
\label{eq:pre_NE1}
\end{equation}

Further we show that Eq.\eqref{eq:pre_NE1} is integrable under certain conditions on parameters $\alpha$ and $\beta$. Exact solutions of Eq.\eqref{eq:pre_NE1} will be obtained in the integrable and non--integrable cases as well.

\section{The Painlev\`{e} test for the extended nonlinear equation.}

Let us consider the application of the Painlev\'{e} test \cite{Weiss,Kudryashov_book,Hone2005} to Eq.\eqref{eq:pre_NE1}. We seek a solution of Eq.\eqref{eq:pre_NE1} in the form \cite{Weiss,Kudryashov_book,Hone2005}
\begin{equation}
v(x,t)=\sum_{j=0}^{\infty} v_{j}\psi^{j-p}, \quad \psi=\psi(x,t)\,.
\label{eq:P_exp}
\end{equation}

Substituting \eqref{eq:P_exp} into Eq.\eqref{eq:pre_NE1} and equating coefficients at the lowest power of $\psi$ we find
\begin{equation}
p=1,\quad v_{0}=-\frac{2\beta}{\nu}\psi_{x}\,.
\label{eq:leading_order}
\end{equation}

Using
\begin{equation}
v(x,t)=-\frac{2\beta\psi_{x}}{\nu\psi}+v_{j}\,\psi^{j-1},
\label{eq:F}
\end{equation}
we find the Fuchs indices
\begin{equation}
j=-1,2,3\,.
\end{equation}
Consequently for integrability of Eq.\eqref{eq:pre_NE1} the functions $\psi,v_{2},v_{3}$ have to be arbitrary in expansion \eqref{eq:P_exp}.

Substituting
\begin{equation}
v(x,t)=-\frac{2\beta\psi_{x}}{\nu\psi}+v_{1}+v_{2}\,\psi+v_{3}\,\psi^{2},
\label{eq:P_exp_short}
\end{equation}
into Eq.\eqref{eq:pre_NE1} and equating coefficients at $\psi$ in $-3,-2$ and $-1$ powers to zero we obtain the system of three equations. From the first equation we find that $v_{1}$ has the form
\begin{equation}
v_{1}=\frac{1}{\nu^{2}\psi_{x}}\left[\nu(\beta\psi_{xx}+\gamma\psi_{t})-(\mu\nu-\alpha\beta)\psi_{x}\right].
\end{equation}

Substituting the value $v_{1}$ into the two other equations we obtain
\begin{equation}
\nu(\nu+\alpha\gamma)\psi_{t}-\alpha(\mu\nu-\alpha\beta)\psi_{x}=0,
\label{eq: pre_correlations_1}
\end{equation}
\begin{equation}
\frac{\partial}{\partial x}(\nu(\nu+\alpha\gamma)\psi_{t}-\alpha(\mu\nu-\alpha\beta)\psi_{x})=0\,.
\label{eq: pre_correlations_2}
\end{equation}

From Eqs.\eqref{eq: pre_correlations_1}, \eqref{eq: pre_correlations_2} we see that Eq.\eqref{eq:pre_NE1} passes the Painlev\`{e} test under the following conditions
\begin{equation}
\nu+\alpha\gamma=0, \quad \mu\nu-\beta\alpha=0\,.
\label{parametrs_correlations}
\end{equation}

Using \eqref{parametrs_correlations} we obtain conditions for $\alpha$ and $\beta$
\begin{equation}
\alpha=-\frac{\nu}{\gamma},\quad \beta=-\gamma\mu\,.
\end{equation}

In this case Eq.\eqref{eq:pre_NE1} can be written as
\begin{equation}
v_{t}-\frac{\nu}{\gamma} v v_{x}-\gamma\mu v_{xxx}-\mu v_{xx}-\nu (v v_{x})_{x}+\gamma v_{xt}=0\,.
\label{eq:pre_NE11}
\end{equation}

However we see that conditions \eqref{parametrs_correlations} on the parameters, which are required for integrability, are not physically realistic. All of the physical parameters $\alpha,\,\beta,\,\gamma,\,\nu,\,\mu$ have to be positive, so the relations (4.10) can never hold.

With the help of the transformations
\begin{equation}
x=\gamma x',\quad t=\frac{\gamma^{2}}{\mu}\,t',\quad v=\frac{\mu}{\nu} v',
\label{eq:integrable_transformations}
\end{equation}
Eq.\eqref{eq:pre_NE11} is reduced to the simple form (the primes are omitted)
\begin{equation}
v_{t} - v v_{x}- v_{xxx} - v_{xx} - (v v_{x})_{x}+v_{xt}=0\,.
\label{eq:NE}
\end{equation}

Let us show that Eq.\eqref{eq:NE} is integrable.  Indeed Eq.\eqref{eq:NE}  can be linearized by the Cole--Hopf transformation.
Substituting
\begin{equation}
v(x,t)=\frac{2\psi_{x}}{\psi},
\label{eq: Cole-Hopf_transformation}
\end{equation}
into Eq.\eqref{eq:NE} we obtain
\begin{equation}
\begin{gathered}
\PDfrac{}{x}\Bigg(\PDfrac{}{x}+1\Bigg)\Bigg\{\frac{\psi_{t}-\psi_{xx}}{\psi}\Bigg\}=0\,.
\label{eq: pre_L_2}
\end{gathered}
\end{equation}

We see that solutions of Eq.\eqref{eq:NE} can be obtained from solutions of the following linear heat equation
\begin{equation}
\begin{gathered}
\psi_{t}-\psi_{xx}=(C_{2}(t)e^{-x}+C_{1}(t))\psi,
\label{general_heat_equation}
\end{gathered}
\end{equation}
where $C_{1}(t)$, $C_{2}(t)$ --- are arbitrary functions of $t$. Solving the Cauchy problem for \eqref{general_heat_equation} and using formulae \eqref{eq: Cole-Hopf_transformation} we obtain solution of the Cauchy problem for Eq.\eqref{eq:NE}.  Without loss of generality we can consider the case $C_{1}(t)=0$. One can remove an arbitrary function $C_{1}(t)$ from Eq.\eqref{general_heat_equation} by scaling the dependent variable $\psi \rightarrow \exp\{\int C_{1} dt\} \psi $.

Let us assume that $C_{2}(t)=e^{t}$. In this case we have the following particular solution of Eq.\eqref{general_heat_equation} in terms of Bessel functions:
\begin{equation}
\psi(x,t)=\exp\left\{\frac{\lambda^{2}+1}{4}\,t-\frac{x}{2}\right\}\Big[C_{3}\,J_{\lambda}(2e^{-(x-t)/2})+
C_{4}\,Y_{\lambda}(2e^{-(x-t)/2})\Big],
\end{equation}
where $\lambda,C_{3},C_{4}$ are arbitrary constants.

If $C_{2}(t)=0$ Eq.\eqref{general_heat_equation} has the following solution
\begin{equation}
\psi(x,t)=\left(\exp\left\{-\lambda^{2}\,t\right\}\left(C_{3}\cos(\lambda x)+C_{4}\sin(\lambda x)\right)+C_{5}\right),
\label{eq: quasy_periodic_solution}
\end{equation}
where $\lambda,C_{3},C_{4},C_{5}$ are arbitrary constants.

One can obtain the transformation for finding solutions of  Eq.\eqref{eq:NE}. Since coefficients $v_{2}$, $v_{3}$ are arbitrary then we can set $v_{2}=v_{3}=0$ in expansion \eqref{eq:P_exp_short}. Substituting into Eq.\eqref{eq:NE} the transformation
\begin{equation}
v=\frac{2\psi_{x}}{\psi}+v_{1},
\label{truncated_expansion}
\end{equation}
we obtain the system of equations
\begin{equation}
\psi_{t}-\psi_{xx}-v_{1}\,\psi_{x}=0,
\label{eq:c11}
\end{equation}
\begin{equation}
\psi_{xx}\Big[\psi_{t}-\psi_{xx}-v_{1}\,\psi_{x}\Big]+\psi_{x}\left\{\psi_{t}-\psi_{xx}-v_{1}\,\psi_{x}+2\,\frac{\partial}{\partial x}\Big[\psi_{t}-\psi_{xx}-v_{1}\,\psi_{x}\Big]\right\}=0,
\label{eq:c21}
\end{equation}
\begin{equation}
\frac{\partial}{\partial x}\left\{\psi_{t}-\psi_{xx}-v_{1}\,\psi_{x}+\frac{\partial}{\partial x}\Big[\psi_{t}-\psi_{xx}-v_{1}\,\psi_{x}\Big]\right\}=0,
\label{eq:c31}
\end{equation}
\begin{equation}
v_{1t} - v_{1}v_{1x}- v_{1xxx} - v_{1xx} - (v_{1}v_{1x})_{x}+v_{1xt}=0\,.
\label{eq:c41}
\end{equation}

It follows from Eqs.\eqref{eq:c11}--\eqref{eq:c41} that if $v_{1}$ is solution of Eq.\eqref{eq:c41} and $\psi$ is found from Eq.\eqref{eq:c11} one can obtain solutions of Eq.\eqref{eq:NE} by formula \eqref{truncated_expansion}.

Assuming in \eqref{truncated_expansion} $v_{1}=\psi(x,t)$ we obtain the formula
\begin{equation}
v=\frac{2\psi_{x}}{\psi}+\psi\,.
\label{truncated_expansion1}
\end{equation}
Using formula \eqref{truncated_expansion1} one can also look for solutions of Eq.\eqref{eq:NE}.
Eq.\eqref{eq:NE} can be presented in the form
\begin{equation}
\left(1+\frac{\partial}{\partial x}\right)\left(v_{t}-\frac{\partial}{\partial x}\left(\frac{\partial}{\partial x}+\frac{v}{2}\right)v \right)=0\,.
\label{eq:NE_transformations2}
\end{equation}
From \eqref{eq:NE_transformations2} one can see that Eq.\eqref{eq:NE}
is a generalized Burgers equation.

We can suggest some hierarchies of nonlinear differential equations with similar properties in the form
\begin{equation}
\sum_{m=1}^{M}\,\sum_{k=1}^{K}A_{mn}\frac{\partial^m}{\partial x^m} \left(v_{t}-\frac{\partial}{\partial x}\left(\frac{\partial}{\partial x}+\frac{v}{2}\right)^k v \right)=0,
\label{eq:NE_transformations5}
\end{equation}
where $A_{mn}$ are the coefficients of Eq.\eqref{eq:NE_transformations5}, $M$ and $K$ are integers.

Note that some classes of the generalized Burgers equation were also considered in \cite{Kudryashov1992}.

\section{Exact traveling wave solutions of equation \eqref{eq:NE}}

Let us find solutions of Eq.\eqref{eq:NE} using the traveling wave variables.
Using transformation $v(x,t)=y(z),\,z=x-C_{0}t$ and integrating once with respect to $z$ we obtain the equation in the form
\begin{equation}
C_{1}-C_{0}\,y-\frac{y^{2}}{2}-y_{zz}-(C_{0}+1)y_{z}-y\,y_{z}=0\,.
\label{eq:rNE}
\end{equation}
where $C_{1}$ is an integration constant.

From Eq.\eqref{eq:rNE} we have
\begin{equation}
\frac{d}{dz}\Big\{-e^{z}y_{z}-e^{z}(y^{2}/2+C_{0}y-C_{1})\Big\}=0\,.
\label{eq:rNE2}
\end{equation}

Integrating once with respect to $z$ we obtain the Riccati equation
\begin{equation}
y_{z}+\frac{y^{2}}{2}+C_{0}y-C_{1}+C_{2}e^{-z}=0\,.
\label{eq:rNE3}
\end{equation}

Using the variable transformations
\begin{equation}
y=v-C_{0},\quad z=2\,z',
\end{equation}

Eq.\eqref{eq:rNE3} can be transformed to the canonical form (the primes are omitted)
\begin{equation}
v_{z}=-v^{2}+(C_{0}^{2}+2C_{1}-2C_{2}e^{-2z})\,.
\label{eq:rNE4}
\end{equation}

Let us show that general solution of Eq.\eqref{eq:rNE4} can be expressed via Bessel functions. Substituting $v=\frac{\Psi_{z}}{\Psi}$ into Eq.\eqref{eq:rNE4} we obtain the following equation
\begin{equation}
\Psi_{zz}+(2C_{2}e^{-2z}-C_{0}^{2}-2C_{1})\Psi=0\,.
\label{eq:rNE5}
\end{equation}

Using the transformation
\begin{equation}
\sqrt{2C_{2}}e^{-z}=\xi,
\end{equation}
from Eq.\eqref{eq:rNE5} we obtain
\begin{equation}
\Psi_{\xi\xi}+\frac{1}{\xi}\Psi_{\xi}+\left(1-\frac{\lambda^{2}}{\xi^{2}}\right)\Psi=0, \quad \lambda^{2}=C_{0}^{2}+2C_{1}\,.
\label{eq:rNE6}
\end{equation}

The general solution of Eq.\eqref{eq:rNE6} has the form
\begin{equation}
\Psi=C_{3}J_{\lambda}(\xi)+C_{4}Y_{\lambda}(\xi)\,.
\label{eq:pre_twgs}
\end{equation}

Using \eqref{eq:pre_twgs} we obtain the general solution of Eq.\eqref{eq:rNE}:
\begin{equation}
y=\frac{2\frac{d}{dz}\Big[C_{3}J_{\lambda}(\sqrt{2C_{2}}e^{-z/2})+C_{4}Y_{\lambda}(\sqrt{2C_{2}}e^{-z/2})\Big]}
{C_{3}J_{\lambda}(\sqrt{2C_{2}}e^{-z/2})+C_{4}Y_{\lambda}(\sqrt{2C_{2}}e^{-z/2})}-C_{0}, \quad \lambda^{2}=C_{0}^{2}+2\,C_{1}\,.
\label{eq:twgs}
\end{equation}
It is clear that solution \eqref{eq:twgs} depends on three arbitrary constants. They are $\lambda$, $C_{2}$ and the ratio $C_{3}/C_{4}$.

\section{Exact solutions of the non--integrable equation.}

Let us consider Eq.\eqref{eq:pre_NE1} in the general case, without imposing  conditions \eqref{parametrs_correlations} on the parameters. Without loss of generality, by scaling, we may assume   $\alpha=1$, $\nu=1$ in \eqref{eq:pre_NE1}. From Eq.\eqref{eq:pre_NE1} we have
\begin{equation}
v_{t}+ v v_{x}+\beta v_{xxx}-\mu v_{xx}-(v v_{x})_{x}+\gamma v_{xt}=0\,.
\label{eq:gen_NE}
\end{equation}

For constructing exact solutions of Eq.\eqref{eq:gen_NE} we use the truncated expansion method in the invariant form \cite{Kudryashov_book,Kudryashov1990,Kudryashov1991,Kudryashov1993}. We search for the solutions of Eq.\eqref{eq:gen_NE} in the form
\begin{equation}
v=\frac{v_{0}}{\chi}+v_{1},
\label{inv_truncation}
\end{equation}
where $\chi=\left(\frac{\psi_{x}}{\psi}-\frac{\psi_{xx}}{2\psi_{x}}\right)^{-1}$. The variable $\chi$ is connected with the Schwarzian derivative $S$ and the variable $C$ by the following relations:
\begin{equation}
\chi_{x}=1+\frac{1}{2}S\,\chi^{2},
\label{e1}
\end{equation}
\begin{equation}
\chi_{t}=-C+C_{x}\chi-\frac{1}{2}(C\,S+C_{xx})\chi^{2}\,.
\label{e2}
\end{equation}
The compatibility condition for equations \eqref{e1} and \eqref{e2} has the form
\begin{equation}
S_{t}+C_{xxx}+2SC_{x}+S_{x}C=0\,.
\label{compatibility_condition}
\end{equation}
Substituting \eqref{inv_truncation} into Eq.\eqref{eq:gen_NE}, using Eqs.\eqref{e1},\eqref{e2} and collecting coefficients at the same powers of $\chi$ we obtain the system of equations for $v_0$, $v_1$, $C$, $S$. Solving this system of equations and using compatibility condition \eqref{compatibility_condition} we find
\begin{equation}
\begin{gathered}
v_{0}=-2\beta,\quad v_{1}=C, \quad C=-\frac{\mu-\beta}{\gamma+1},\\
S=A+B\,e^{x-C\,t},
\label{parameters1}
\end{gathered}
\end{equation}
where $A,B$ are an arbitrary constants.

Solving system of Eqs.\eqref{e1}, \eqref{e2} we find the invariant variable $\chi$
\begin{equation}
\begin{gathered}
\chi=\frac{\Psi}{\Psi_{x}},\quad \Psi=C_{1}\,J_{\sqrt{-2\,A}}\left(\sqrt{2B}\,e^{\frac{x-Ct}{2}}\right)+C_{2}\,Y_{\sqrt{-2\,A}}
\left(\sqrt{2B}\,e^{\frac{x-Ct}{2}}\right)\,.
\label{chi_solution}
\end{gathered}
\end{equation}
Using \eqref{parameters1}, \eqref{chi_solution} we obtain solutions of Eq.\eqref{eq:gen_NE}  in the form
\begin{equation}
\begin{gathered}
v=-2\,\beta\frac{\Psi_{x}}{\Psi}+C, \quad \Psi=C_{1}\,J_{\sqrt{-2\,A}}\left(\sqrt{2B}\,e^{\frac{x-Ct}{2}}\right)+C_{2}\,Y_{\sqrt{-2\,A}}
\left(\sqrt{2B}\,e^{\frac{x-Ct}{2}}\right)\,.
\label{genNE_solution}
\end{gathered}
\end{equation}
It is clear that solution \eqref{genNE_solution} contains three arbitrary constants and this solution is the traveling wave solution. However Eq.\eqref{eq:gen_NE}  does not possess the Painlev\'e test in the general case.

In the case $A=0$ we find the solitary wave solution of Eq.\eqref{eq:gen_NE} in the from
\begin{equation}
\begin{gathered}
v=-2\,\beta\frac{\Psi_{x}}{\Psi}+C, \quad \Psi=C_{1}\,e^{\sqrt{\frac{1}{2}}(x-C\,t)}+C_{2}\,e^{-\sqrt{\frac{1}{2}}(x-C\,t)},
\label{genNE_solution1}
\end{gathered}
\end{equation}
with two arbitrary constants. We note that solution \eqref{genNE_solution1} is the kink.

Consider Eq.\eqref{eq:pre_NE1} in the form
\begin{equation}
v_{t}+\alpha v v_{x}+\beta v_{xxx}-\mu v_{xx}-\nu (v v_{x})_{x}+\gamma v_{xt}=0\,.
\label{eq:NE_transformations}
\end{equation}

Using transformations
\begin{equation}
\xi=x+\frac{\alpha\mu}{\alpha\gamma+\nu}t,\quad \tau=t, \quad w=v+\frac{\mu}{\alpha\gamma+\nu},
\label{eq:transf}
\end{equation}

Eq.\eqref{eq:NE_transformations} for $\alpha\gamma+\nu\neq0$ is transformed to the following:
\begin{equation}
w_{\tau}+\alpha w w_{\xi}+\beta w_{\xi\xi\xi}-\nu (w w_{\xi})_{\xi}+\gamma w_{\xi\tau}=0\,.
\label{eq:Kawahara}
\end{equation}

Eq.\eqref{eq:Kawahara} was first obtained in \cite{Kawahara} for the description of the ion-acoustic waves in plasma. Exact solutions of Eq.\eqref{eq:Kawahara} were obtained in \cite{Isidore}. Let us note that exact solutions \eqref{genNE_solution}, \eqref{genNE_solution1} can be found from the solutions obtained in \cite{Isidore} with the help of transformations \eqref{eq:transf}.

In the case of $\alpha\gamma+\nu=0$  Eq.\eqref{eq:NE_transformations} is invariant under the Galilean group
\begin{equation}
\xi=x+a\,t,\quad \tau=t+a, \quad w=v-\frac{a\gamma}{\nu},
\label{eq:group}
\end{equation}
where $a$ is the group parameter. So, Eq.\eqref{eq:NE_transformations} at condition $\alpha\gamma+\nu=0$ admits the Galilean symmetry \eqref{eq:group}. The condition $\alpha\gamma+\nu=0$ includes the integrable case of Eq.\eqref{eq:NE_transformations}. We considered this case in Section 4.

If $\alpha\gamma+\nu=0$  with the help of transformations \eqref{eq:integrable_transformations} Eq.\eqref{eq:NE_transformations} can be transformed to the following form (the primes are omitted)
\begin{equation}
v_{t}-v\,v_{x}+\tilde{\beta}v_{xxx}-v_{xx}-(v\,v_{x})_{x}+v_{xt}=0{\color{red},}
\label{eq:NE_Galilean}
\end{equation}
where $\tilde{\beta}=\frac{\beta}{\mu\gamma}$.

Using symmetry \eqref{eq:group} and taking into account transformations \eqref{eq:integrable_transformations} and condition $\alpha\gamma+\nu=0$ we find the following variables
\begin{equation}
v=w(\theta)-t,\quad \theta=x-\frac{t^{2}}{2}\,.
\label{eq:Galilean_variables}
\end{equation}

Introducing in \eqref{eq:NE_Galilean} variables \eqref{eq:Galilean_variables} and integrating once with respect to $\theta$ we obtain
\begin{equation}
\tilde{\beta} w_{\theta\theta}-w_{\theta}-w w_{\theta}-\frac{1}{2}w^{2}=\theta+C_{1},
\label{eq:NE_transformations_reduction}
\end{equation}
where $C_{1}$ is an integration constant.

One can show that Eq.\eqref{eq:NE_transformations_reduction} passes the Painlev\'{e} test only at $\tilde{\beta}=-1$ and $\tilde{\beta}=0$. The case of $\tilde{\beta}=0$ is trivial. The case of $\tilde{\beta}=-1$ correspond to the integrable variant of Eq.\eqref{eq:NE_transformations}. Indeed at $\tilde{\beta}=-1$ Eq.\eqref{eq:NE_transformations_reduction} can be presented in the form
\begin{equation}
q_{\theta}+q=-\theta-C_{1}, \quad w_{\theta}+\frac{w^{2}}{2}=q, \quad q=q(\theta)\,.
\label{eq:integrable_Galilean_reduction}
\end{equation}

We see that $w$ can be found from the Riccati equation. Right--hand side of this Riccati equation is the solution of the linear first order differential equation.

If $\tilde{\beta}\neq-1,0$ then Eq.\eqref{eq:NE_transformations_reduction} does not pass the Painlev\'{e} test. In this case Eq.\eqref{eq:NE_transformations_reduction} admits non--meromorphic solutions. One can construct the psi--series containing logarithmic terms for the local solution of Eq.\eqref{eq:NE_transformations_reduction}.

\section{Numerical simulation of nonlinear wave processes}

Let us study the nonlinear wave processes governed by Eq.\eqref{eq:pre_NE1} numerically. We consider the boundary--value problem for Eq.\eqref{eq:pre_NE1} with periodic boundary conditions and initial condition. With this aim we use the integrating factor with the fourth-order Runge–-Kutta approximation method (IFRK4) introduced in \cite{Cox,Kassam}.

We rewrite Eq.\eqref{eq:pre_NE1} in the form
\begin{equation}
v_{t}=-\gamma v_{xt}+L(v)+N(v),
\label{NE_bft}
\end{equation}
where $L$ is the linear differential operator from Eq.\eqref{eq:pre_NE1} and $N$ is the nonlinear one. The boundary and initial conditions are the following
\begin{equation}
v(x,t)=v(x+H,t), \quad v(x,0)=v_{0}(x),
\label{NE_bound_conditions}
\end{equation}
where $H$ is the period.

We discretize the spatial part of Eq.\eqref{NE_bft} using the Fourier transform. As the result we have the following problem for the system of ordinary differential equations
\begin{equation}
(1+i\,k\,\gamma) v_{t}=\hat{L}(v)+\hat{N}(v), \quad v(x,0)=v_{0}(x)\,.
\label{NE_aft_1}
\end{equation}

We divide  Eq.\eqref{NE_aft_1} by $1+i\,k\,\gamma$ and obtain the following problem
\begin{equation}
v_{t}=\tilde{L}(v)+\tilde{N}(v), \quad v(x,0)=v_{0}(x),
\label{NE_aft_2}
\end{equation}
where $\tilde{L}=\hat{L}/(1+i\,k\,\gamma)$, $\tilde{N}=\hat{N}/(1+i\,k\,\gamma)$.

Applying the integrating factor in Eq.\eqref{NE_aft_2}
\begin{equation}
v=e^{-\tilde{L}t}\,w,
\end{equation}
we get
\begin{equation}
w_{t}=e^{-\tilde{L}t}\tilde{N}(e^{\tilde{L}t}w)\,.
\label{NE_aft_2b}
\end{equation}
We solve this system of ordinary differential equation using the fourth-order Runge--Kutta method.

\begin{figure}[h] 
  \centering      
 \includegraphics[width=0.49\textwidth]{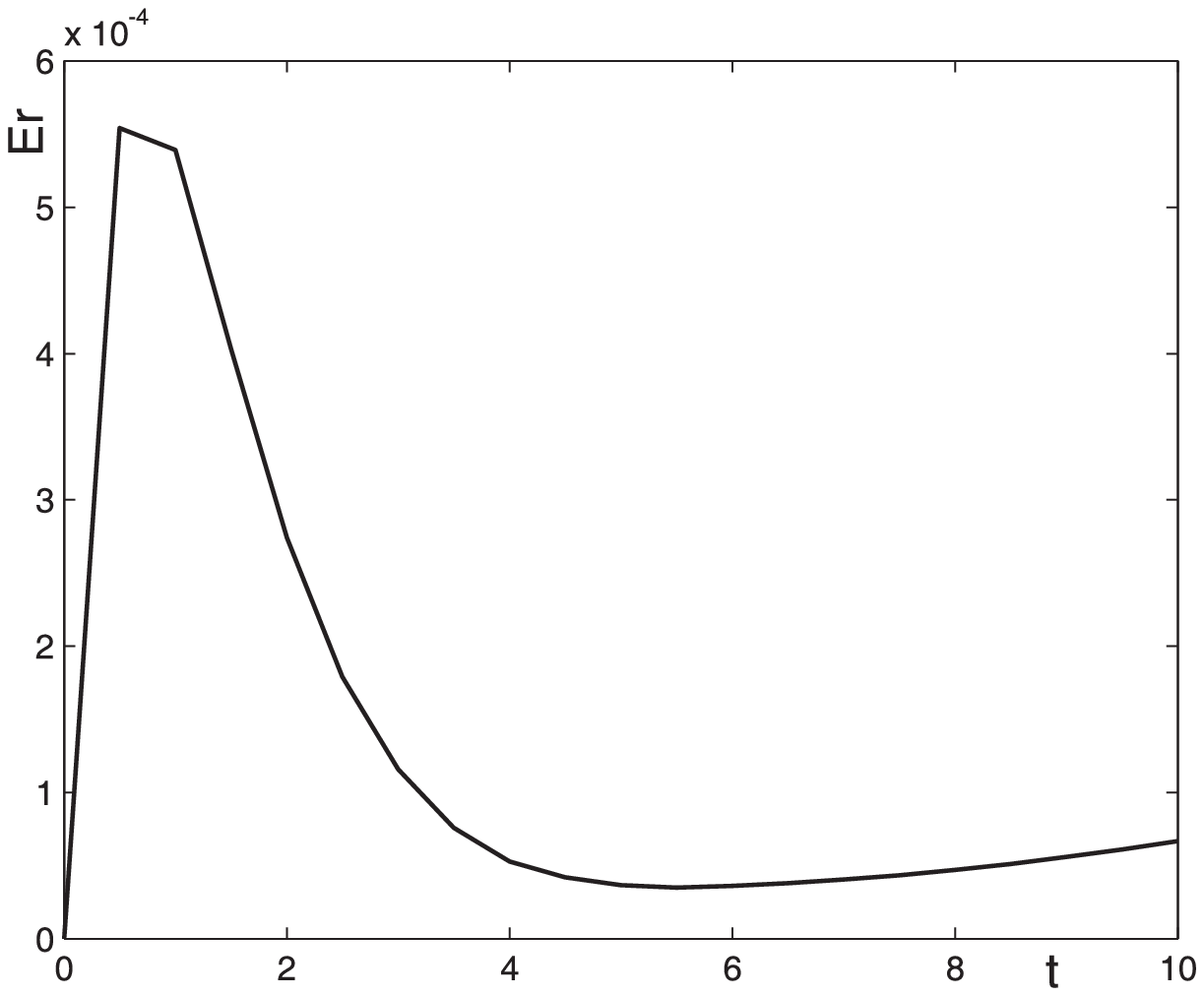}
 \includegraphics[width=0.49\textwidth]{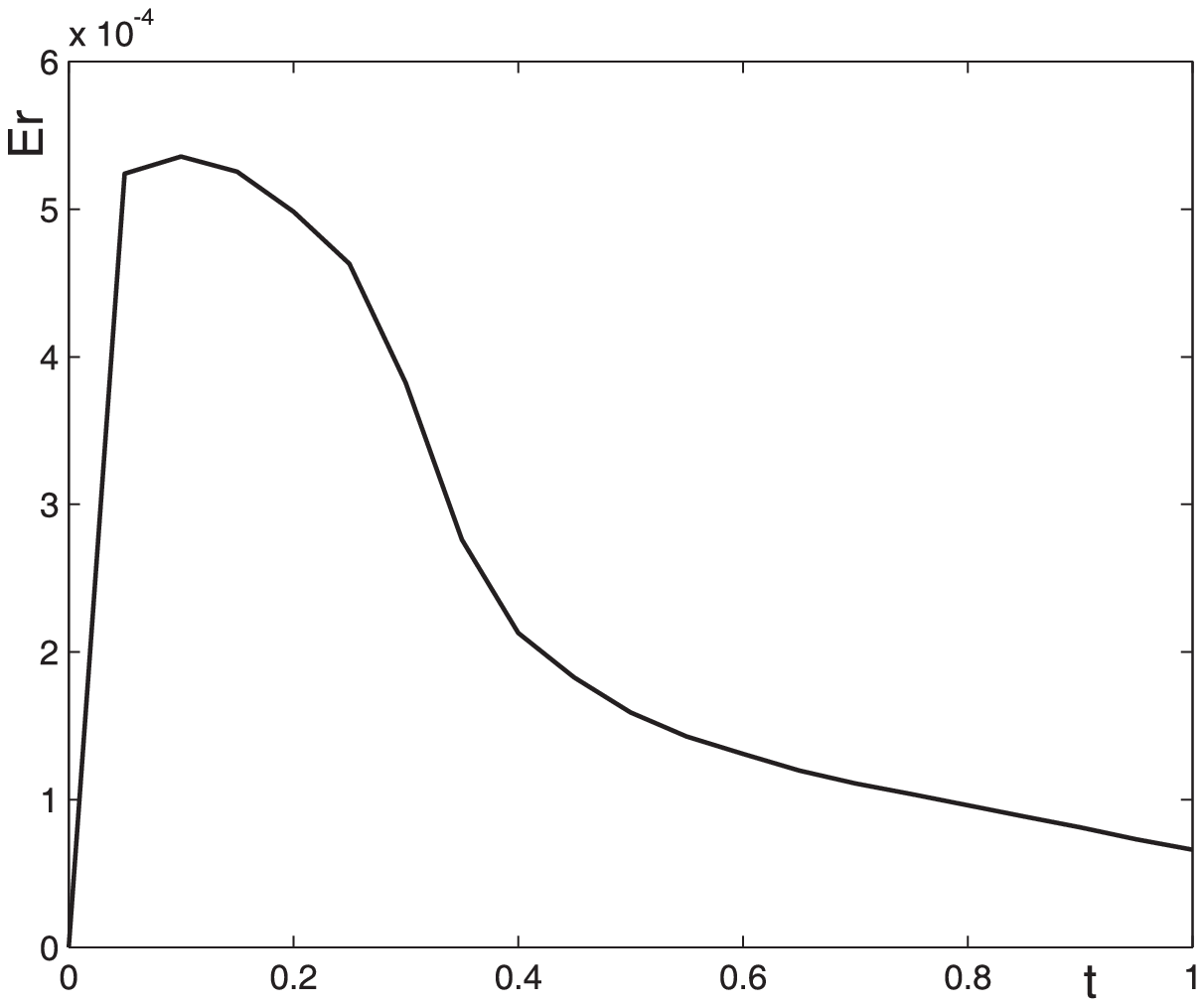}
    \caption{Average error for solutions \eqref{eq: quasy_periodic_solution} and \eqref{genNE_solution1} respectively.}
  \label{fig:1}
\end{figure}

\begin{figure}[h] 
  \centering      
 \subfigure[]{\includegraphics[width=0.49\textwidth]{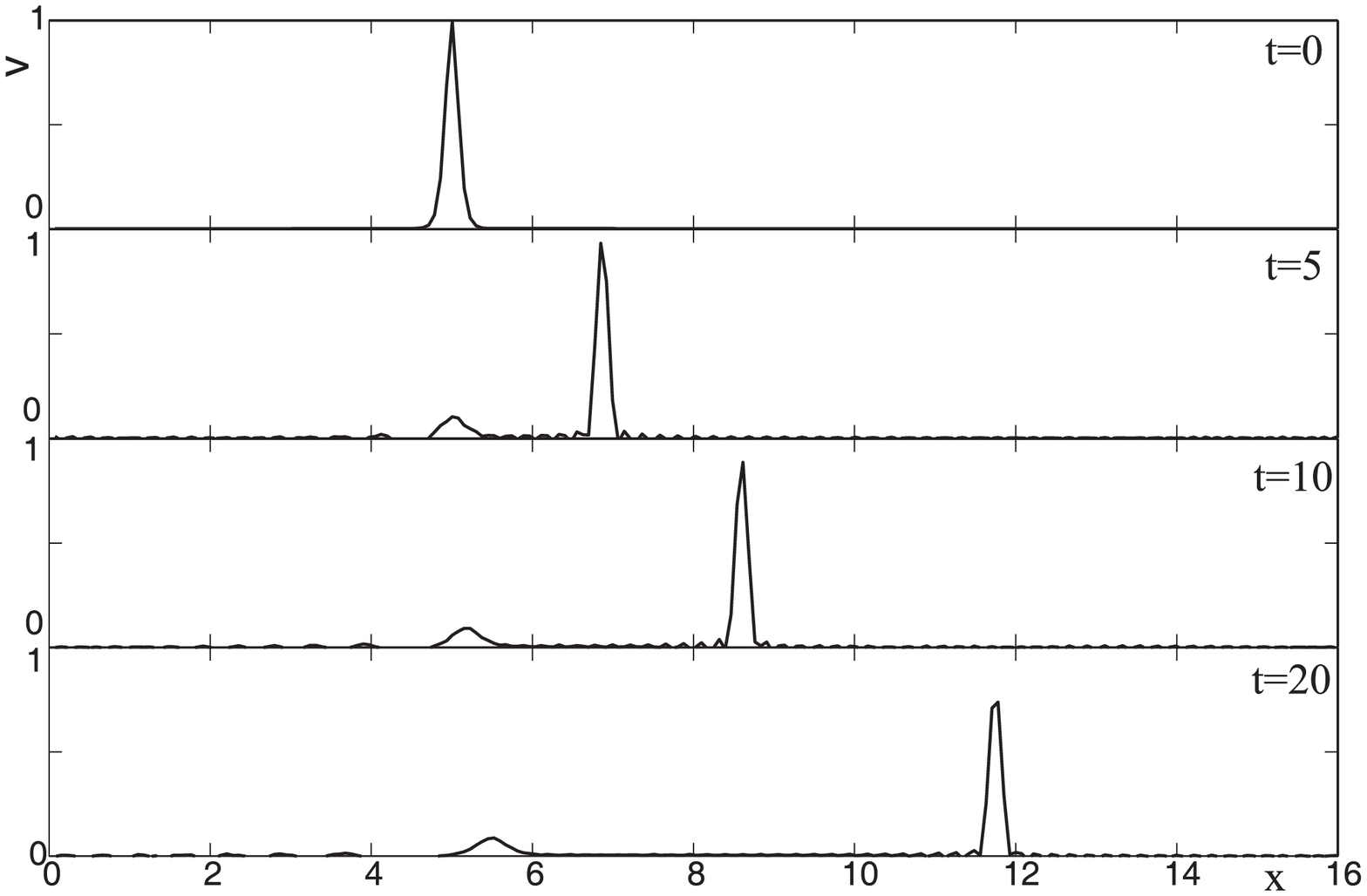}}
 \subfigure[]{\includegraphics[width=0.49\textwidth]{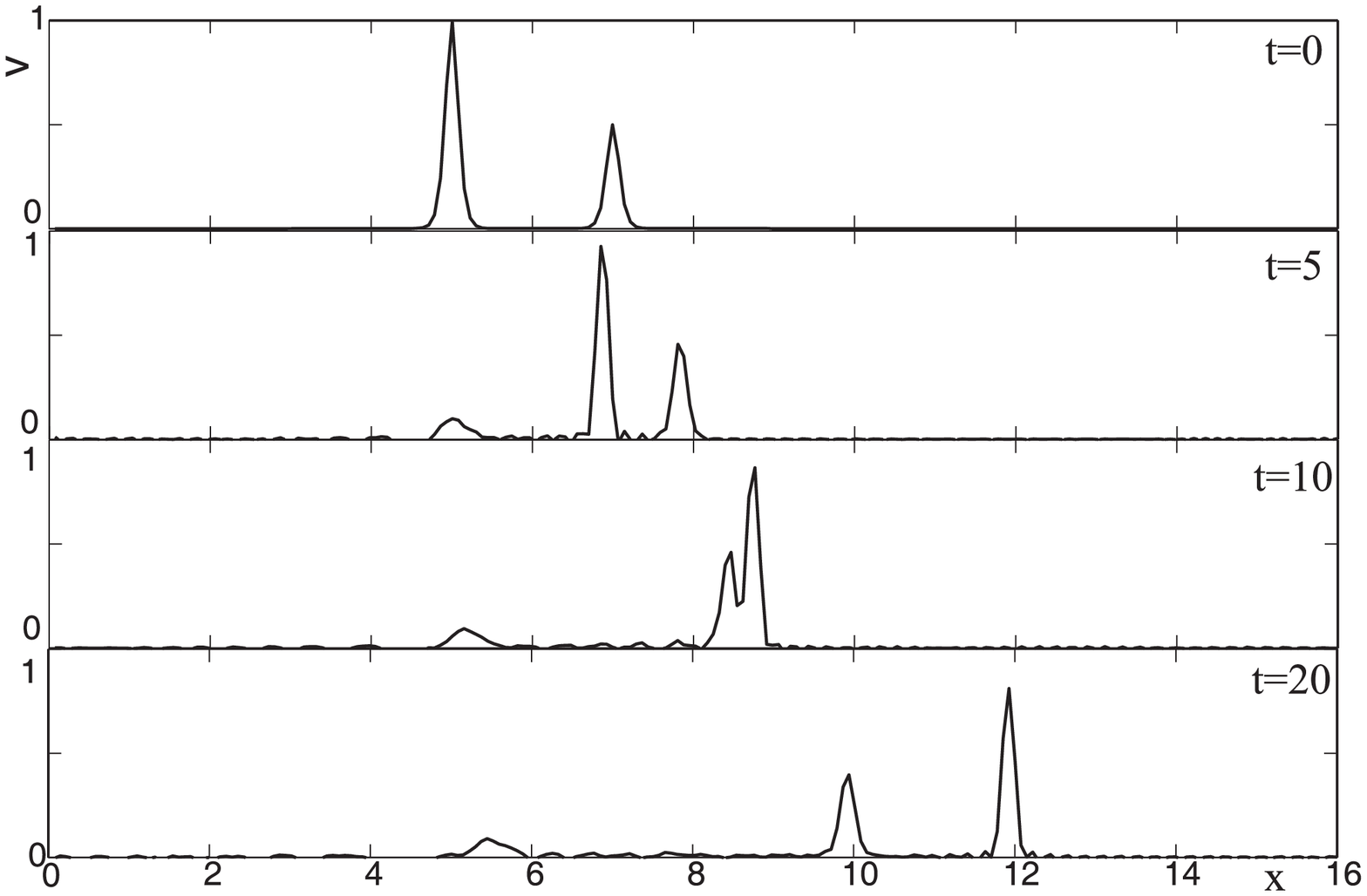}}
    \caption{Evolution of a solitary wave (a) and interaction of two solitary waves (b).}
  \label{fig:2}
\end{figure}

To verify our numerical calculations we use exact solutions \eqref{eq: quasy_periodic_solution} and \eqref{genNE_solution1}.

Let us define the following average error
\begin{equation}
Er=\frac{1}{M}\sum\limits_{i=1}^{M}|v_{exact}^{i}-v_{num}^{i}|,
\label{eq:num_err}
\end{equation}
where $M$ is the number of approximation points on the $x$ axis. In all calculations we use $M=256$. We use solution  \eqref{eq: quasy_periodic_solution} at $t=0$ as initial condition for problem \eqref{NE_aft_2}. We note that solution \eqref{genNE_solution1} is not periodic. We mirror-reflect this solution about a certain point and use the superposition of exact and reflected solutions as initial condition. We demonstrate evolution of average error \eqref{eq:num_err} for solution  \eqref{eq: quasy_periodic_solution} at $\alpha=\beta=-\mu=-\nu=-\gamma=-1$ on Fig.\ref{fig:1}. We also demonstrate evolution of average error \eqref{eq:num_err} for  the superposition of exact solution \eqref{genNE_solution1} and reflected solution \eqref{genNE_solution1} at $\alpha=10,\,\beta=\nu=1,\mu=3,\gamma=0.01$ on Fig.\ref{fig:1}. From Fig.\ref{fig:1} we see that numerical solutions and exact solutions are in close agreement.

\begin{figure}[h] 
  \centering      
 \subfigure[]{ \includegraphics[height=5cm,width=0.48\textwidth]{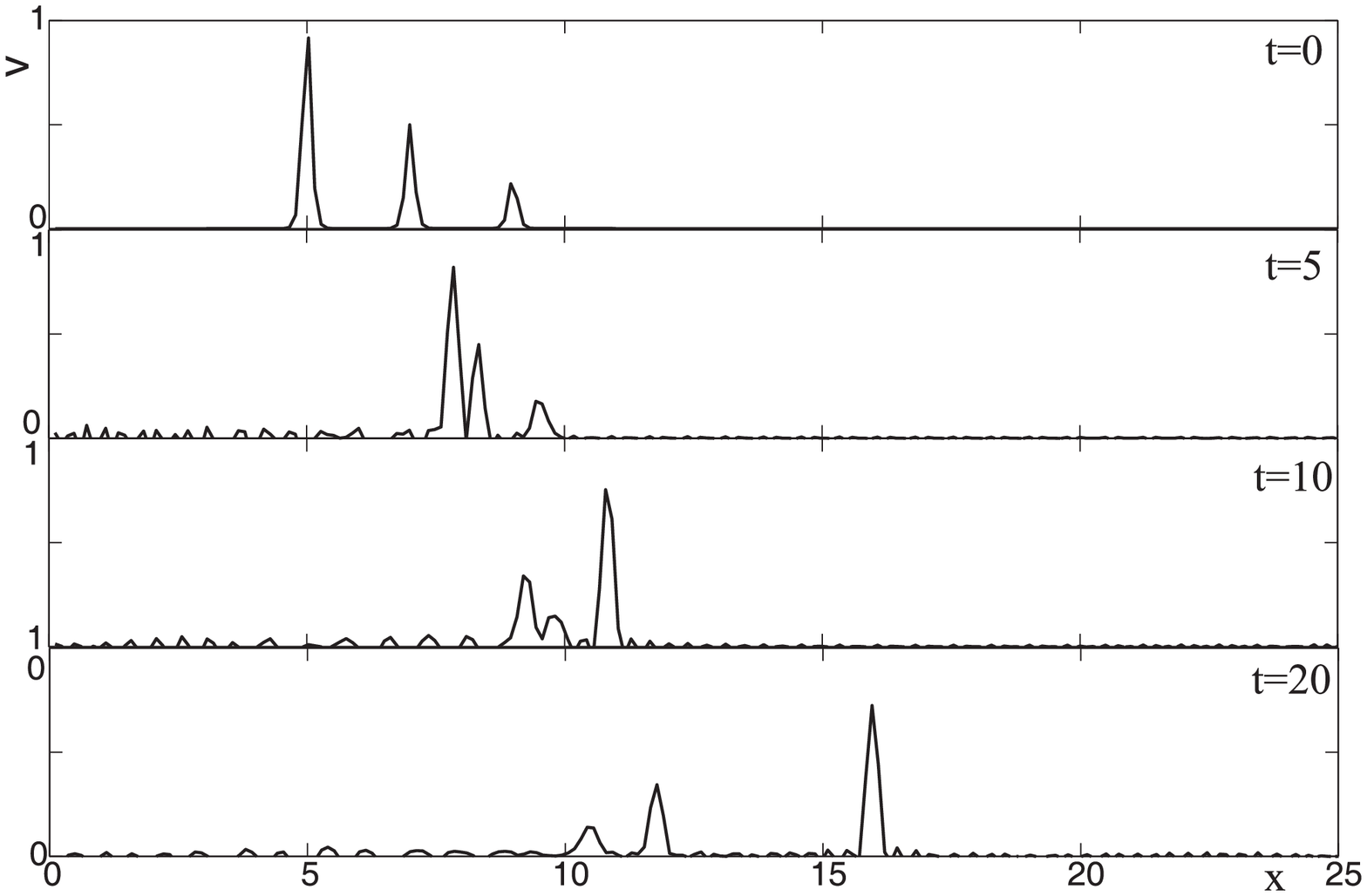}}
 \subfigure[]{ \includegraphics[height=5cm,width=0.48\textwidth]{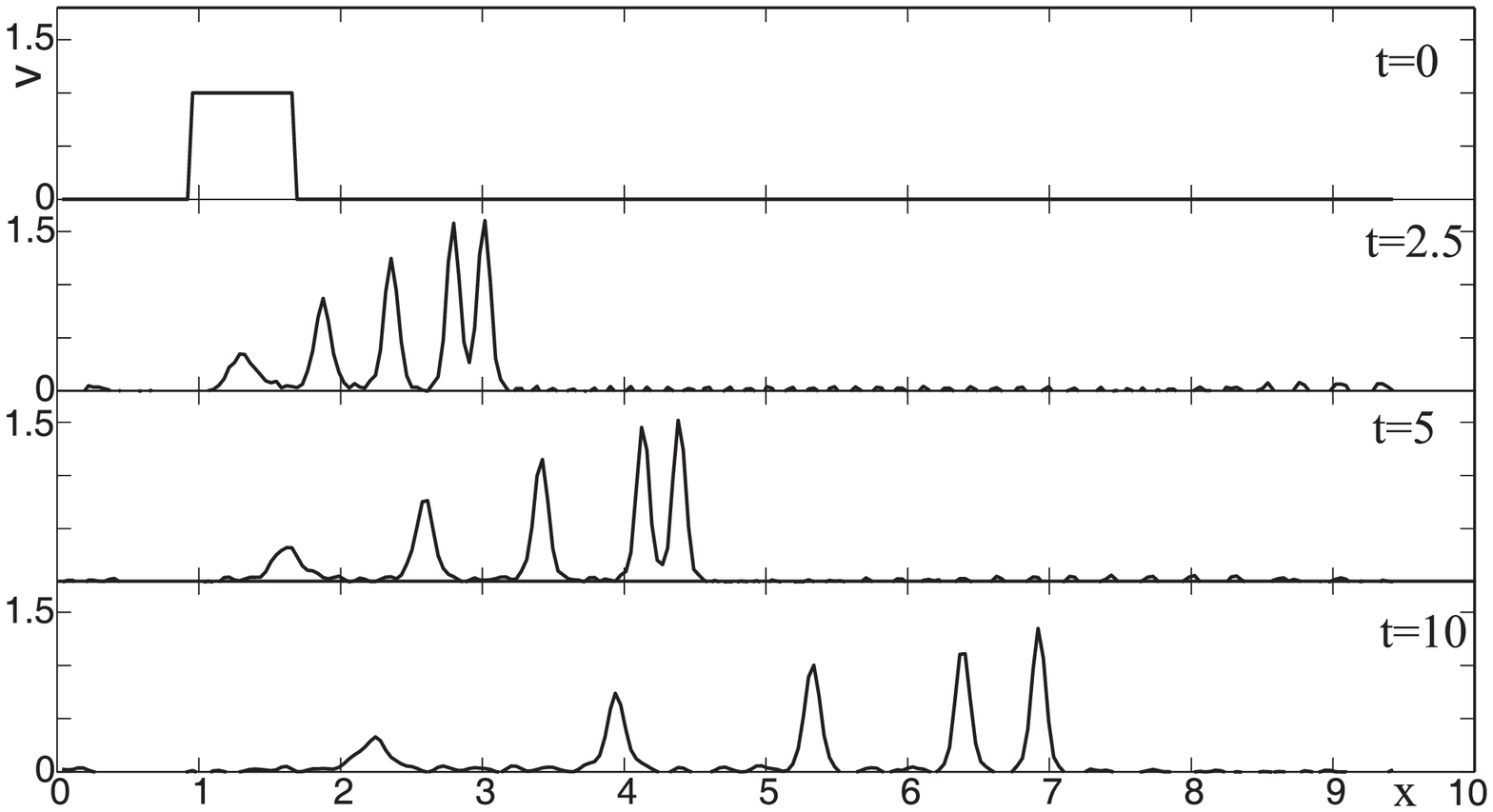}}
    \caption{Interaction of three the solitary wave (a) and evolution of an impulse (b).}
  \label{fig:3}
\end{figure}

We studied the evolution of the various initial profiles using the above described numerical approach. We use the following values of parameters of Eq. \eqref{eq:pre_NE1} for performing numerical calculations: $\alpha=1,\quad \beta=0.0004,\quad \mu=0.0001,\quad \nu=0.00001,\quad \gamma=0.00001$.

On Fig.\ref{fig:2}a the evolution of solitary wave with initial condition $v(x,0)=\mbox{sech}^{2}\{\frac{19(x-5)}{2}\}$ is presented. From Fig.\ref{fig:2}a we see that solitary wave propagates without dramatically changing of its shape. However we see that the oscillating tail is generated behind the solitary wave.

Fig.\ref{fig:2}b shows the process of interaction of two solitary waves. From Fig.\ref{fig:2}b we see that the waves interact with each other and then propagate with a phase shift. The shape of the waves remain the same. However we observe the oscillating tail behind the waves. There is the slight change in the amplitude of the solitary waves as well. We see that the solitary waves do not interact elastically as solitons.

Fig.\ref{fig:3}a illustrates interaction of the three solitary waves. We observe that three solitary waves interact similarly to two solitary waves. We see the change in the amplitude of the waves and the oscillating tail behind the waves again.

On Fig.\ref{fig:3}b we see the evolution of the rectangular impulse. We observe that the rectangular impulse breaks up into several solitary waves changing shape. These solitary waves propagate without shape changing.

\section{Conclusion}
Let us summarize the results of this work. We have considered the long nonlinear waves in a liquid with gas bubbles and  have taken into account the liquid viscosity, the surface tension and the interphase heat exchange in the nearly isothermal approximation. We have derived the nonlinear differential equation for the long weakly nonlinear waves in the bubbly liquid  (Eq.\eqref{eq:pre_NE1}).  This nonlinear differential equation is the extension of another equation for the description of the nonlinear waves in the bubble-liquid mixture with the additional dispersive and dissipative terms of second order with respect to the small parameter. Eq.\eqref{eq:pre_NE1} under some conditions on parameters is equivalent to the equation obtained in \cite{Kawahara} for the ion-acoustic waves in plasma.

We have shown that \eqref{eq:pre_NE1} is integrable under the additional conditions on the parameters of the equation. The transformation linearization of the equation were studied.
However conditions for integrability of Eq.\eqref{eq:pre_NE1} lead to unphysical values of parameters. We have obtained the exact solutions of Eq.\eqref{eq:pre_NE1} in the integrable and non--integrable cases. We have investigated numerically the nonlinear wave process described by Eq.\eqref{eq:pre_NE1} as well.

\section{Acknowledgment}
The authors are grateful to the anonymous referees for helpful suggestions.

This research was partially supported by Federal Target Programm
Research and Scientific-Pedagogical Personnel of Innovation
in Russian Federation on 2009-2013, by RFBR grant 11--01--00798--a
and "Researches and developments in priority directions of development of a scientifically-technological complex of Russia on 2007-2013".

\end{document}